\documentclass[showpacs,a4paper, oneside, aps, preprint,amsmath,amssymb,nofootinbib]{revtex4}
\usepackage{times}
\usepackage[T1]{fontenc}
\usepackage[latin1]{inputenc}
\usepackage{color}
\usepackage{graphicx}
\usepackage{amsfonts}
\usepackage{epsfig}

\makeatletter

\newtheorem{theorem}{Theorem}

\makeatother

\begin{document}		
\title{Poincar\'{e}'s Observation and the Origin of Tsallis Generalized Canonical
Distributions}

\author{C. Vignat{*} and A. Plastino{*}{*}}

\address{{*}L.I.S. Grenoble, BP 46, 38402 St. Martin d'H\`{e}res cedex, France,
vignat@lis.inpg.fr }

\address{{*}{*}Facultad de Ciencias Exactas, National University La Plata and IFLP-CONICET, C. C. 727, 1900 La
Plata, Argentina}


\begin{abstract}
In this paper, we present some geometric properties of the maximum
entropy (MaxEnt) Tsallis-distributions under energy constraint. In
the case $q>1$, these distributions are proved to be marginals of
uniform distributions on the sphere; in the case $q<1$, they can
be constructed as conditional distributions of a Cauchy law built from the same
uniform distribution on the sphere using a gnomonic projection. 
As such, these distributions reveal the
relevance of using Tsallis distributions in the microcanonical
setup: an example of such application is given in the case of the
ideal gas.
\end{abstract}
\maketitle

\section{Introduction}

Nonextensive thermostatistics is a very active field nowadays,
based on the concept of Tsallis' information measure (or
entropy)~\cite{gellmann}.  The concomitant  maximum (Tsallis)
entropy distributions play a role for the description of
non-extensive systems akin of that of Gaussian ones for extensive
systems. By Tsallis distributions, we mean the following
$n-$variate probability densities:
\begin{itemize}
\item if $\frac{n}{n+2}<q<1$, with $\beta=\frac{1}{2q-n\left(1-q\right)}$
and $A_{q}=\frac{\Gamma\left(\frac{1}{1-q}\right)\left(\beta\left(1-q\right)\right)^{n/2}}{\Gamma\left(\frac{1}{1-q}-\frac{n}{2}\right)\pi^{n/2}\vert C\vert^{1/2}}$\[
f_{X}\left(X\right)=A_{q}\left(1-\beta\left(q-1\right)x^{T}C^{-1}x\right)^{\frac{1}{q-1}}\]

\item if $q>1$ or $q<0$, with $\beta=\frac{1}{2q-n\left(1-q\right)}$
and $A_{q}=\frac{\Gamma\left(\frac{q}{q-1}+\frac{n}{2}\right)\left(\beta\left(q-1\right)\right)^{n/2}}{\Gamma\left(\frac{q}{q-1}\right)\pi^{n/2}\vert C\vert^{1/2}}$\[
f_{X}\left(X\right)=A_{q}\left(1-\beta\left(q-1\right)x^{T}C^{-1}x\right)_{+}^{\frac{1}{q-1}}\]
where $x_{+}$ denotes $\max\left(x,0\right)$.
\end{itemize}
These distributions are solutions of the following maximization problem:\[
\max_{f}\frac{1}{1-q}\int_{\mathbb{R}^{\mathnormal{n}}}f^{q}\text{ such that }\int_{\mathbb{R}^{\mathnormal{n}}}xx^{T}f\left(x\right)dx=C. \]
In this sense, they are the counterparts of the Gaussian distribution with covariance matrix $C$ which is solution of 
\[
\max_{f} \left( -\int_{\mathbb{R}^{\mathnormal{n}}}f\log f \right) \text{ such that }\int_{\mathbb{R}^{\mathnormal{n}}}xx^{T}f\left(x\right)dx=C\]

Gaussian distributions exhibit universal properties that allow to derive important statistical results such as the central limit theorem or the entropy power inequality. Some of these results have a ''Tsallis-counterpart" \cite{gellmann}. A geometric approach like the one advanced here is interesting
since it sheds  new light on these properties.

\section{Geometric Approach}

\subsection{stochastic properties}

In this section, we propose a geometric characterization of Tsallis
distributions based on their stochastic properties. Thus we begin
by a detailed stochastic study of Tsallis random vectors. We denote
as \[
X\sim T_{q,C}\]
a Tsallis vector with nonextensivity parameter $q$ and covariance matrix
$C$.

\subsubsection{case $\frac{n}{n+2}<q<1$}

If $X\sim T_{q,C}$, then a stochastic representation of $X$ writes\[
X=\sqrt{m-2}C^{1/2}\frac{N}{\sqrt{a}}\]
where $a$ is a chi-square distributed random variable with $m=\frac{2}{1-q}-n$
degrees of freedom, independent of the Gaussian random vector $N$
with unit covariance matrix. It is understood that $m$ may be non-integer.

\subsubsection{duality result}

If $X\sim T_{q,C}$ with $\frac{n}{n+2}<q<1$ then the following random vector\[
Y=\frac{X}{\sqrt{1+\frac{1}{m-2}X^{T}C^{-1}X}}\]
verifies\[
Y\sim T_{q*,C*}\]
with\[
C*=C\frac{m-2}{m+n}, \frac{1}{q*-1}=\frac{1}{1-q}-\frac{n}{2}-1\]

\subsubsection{case $q>1$}

From the duality result, we deduce a stochastic representation for
$Y\sim T_{q*,C}$ with $q*>1$ as follows\begin{equation}
Y=\sqrt{m+n} \, C^{1/2}\frac{N}{\sqrt{a+N^{T}N}}\label{eq:stochasticY}\end{equation}
where $a$ is a chi-square random variable with $m=\frac{2q*}{q*-1}$
degrees of freedom, independent of the Gaussian random vector $N$
with unit covariance matrix.

\subsection{Geometric representation}

The preceding stochastic representations allow to derive easily a
geometric construction of Tsallis random vectors as follows.

\subsubsection{case $q>1$}

An $n-$variate random vector $U$ is uniformly distributed on the
ellipsoid $\mathcal{E}_{\mathnormal{C,n}}=\left\{
\mathnormal{Z}\in\mathbb{R}^{\mathnormal{n}} \vert \mathnormal{Z}^{\mathnormal{T}}\mathnormal{C}\mathnormal{Z}= \mathnormal{1} \right\} $ if and only if it
writes\[ U=\frac{C^{-1/2}N}{\sqrt{N^{T}N}}\] where $N$ is a
Gaussian vector with unit covariance matrix. Comparing this
stochastic representation with the Tsallis vector's one
(\ref{eq:stochasticY}), the following result can be proved:

\begin{theorem}
\label{thm1}
If $Y\sim T_{q*,C}$ with $q*>1$ or $q*<0$ then $Y$ is the $n-$variate marginal
vector of an $(m+n)-$variate random vector $U$ uniformly distributed on the ellipsoid
$\mathcal{E}_{\mathnormal{C^{-1},m+n}}$ with $m=\frac{2q*}{q*-1}$.
\end{theorem}

We remark that, according to this characterization, the $(m+n-1)-$variate marginal random vector is the only one distributed according to a Tsallis law with negative non-extensivity index (namely $q*=-1$); moreover, the $(m+n-2)-$variate marginal is uniform inside a $(m+n-2)-$ dimensional ellipsoid and corresponds to a non-extensivity index $q*=+\infty$. All other lower dimension marginals have finite and positive index $q*>1$.

\subsubsection{case $q<1$}

For the sake of simplicity, we address here the uncorrelated case $C=I_{n}.$ We
note that this is not a loss of generality, since\[
X\sim T_{q,C}\Rightarrow C^{-1/2}X\sim T_{q,I}.\]

\begin{theorem}
Assume point $P$ is uniformly distributed on the sphere $\mathcal{E}_{\mathnormal{I_{n},n}}$; consider the intersection $M$ of vector $OP$ with the hyperplane $\mathcal{H} = \left \{\mathnormal{Z} \in \mathbb{R}^{\mathnormal{n}}\vert \mathnormal{Z}_{n}=1 \right \} $. \footnote{ point M is called the gnomonic projection of point P}
Then point $M$ follows a $(n-1)-$ variate Cauchy  distribution in $\mathcal{H}$:
\[
f_{M}(y_{1},\dots,y_{n-1}) \propto (1+\sum_{i=1}^{n-1}y_{i}^{2})^{-\frac{n}{2}}
.\]

Moreover, any distribution of point $M$ conditioned on variables $y_{k+1},\dots,y_{n-1}$ follows a $k-$ variate Tsallis distribution with $m=n-k$ degrees of freedom \cite{johnson}:
\[
f_{M}(y_{1},\dots , y_{k} \vert y_{k+1},\dots y_{n-1} ) \propto
(1+\lambda \sum_{i=1}^{k} y_{i}^{2})^{-\frac{n}{2}}
\]
\end{theorem}

In Fig. 1 below, the $n=3$ dimensional sphere is depicted, with the Cauchy distribution of point $M$: the white curve represents, up to a constant, the one dimensional conditional density $f_{M}(y_{1} \vert y_{2})$ which is Tsallis distributed with $m=2$ degrees of freedom ($q=\frac{1}{3}$).

\begin{center}
\begin{figure}
\includegraphics[scale=0.7]{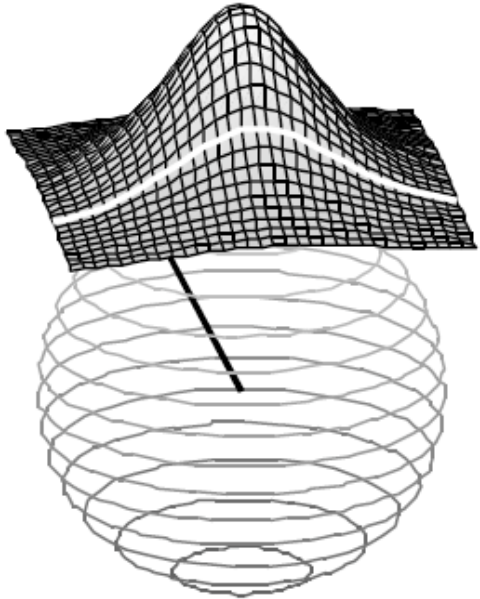}
\caption{ the $n=3$ dimensional sphere, the Cauchy distribution and one of its conditional laws, the Tsallis $q=\frac{1}{3}$ distribution}
\end{figure}
\end{center}

\section{Comments on some particular cases}

In this section, we comment on the geometric results obtained.

\subsection{the case $q>1$ and Poincar\'{e}'s observation}

The characterization of Tsallis random vectors - with $q>1$ and
identity covariance matrix - as marginals of vectors uniformly
distributed on a sphere can be related to a result attributed to
Poincar\'{e} called ''Poincar\'{e}'s Observation'' \cite{Poincare} (for a discussion on this paternity, see \cite{diaconis}).

\begin{theorem}
If $X=\left(x_{1},\dots,x_{p}\right)$ is uniformly distributed on
the sphere $S_{p}$ with radius $\sqrt{p}$, then for fixed
$n<+\infty$,
\texttt{\[\lim_{p\rightarrow\infty}\Pr\left(\cap_{i=1}^{n}\left(a_{i}\le
x_{i}\le
b_{i}\right)\right)
=\int_{a_{1}}^{b_{1}}
\frac{e^{-\frac{x^{2}}{2}}}{\sqrt{2\pi}}dx\dots
\int_{a_{n}}^{b_{n}}\frac{e^{-\frac{x^{2}}{2}}}{\sqrt{2\pi}}dx\]
}
\end{theorem}
This result shows that a Gaussian (Tsallis $q=1$ maximizer) random
vector with fixed dimension $n$ can be considered as the marginal
vector of an infinite-dimensional vector uniformly distributed on
the sphere $S_{\infty}$. The results obtained above demonstrate
that, in contrast, a Tsallis $q>1$ random vector of dimension $n$
can be viewed as the marginal of a finite-dimensional vector
uniformly distributed on the sphere $S_{p}$ with\[
p=\frac{2q}{q-1}+n.\] As $q\rightarrow1^{+},$
$p\rightarrow+\infty$ and we recover Poincar\'e's observation.

\subsection{the case $m=1,q<1$}

The case $m=1$ and $q<1$ corresponds to Cauchy-Lorentz
distributions, which are not {\it stricto sensu} Tsallis
distributions since they have no finite covariance matrix.
However, they appear in many relevant physical situations
\cite{abe}. They are defined as
\[f_{X}\left(X\right)
=\frac{\Gamma\left(\frac{n+1}{2}\right)}{\pi^{\frac{n+1}{2}}}
\left(1+X^{T}X\right)^{-\frac{n+1}{2}}\] and have for stochastic
representation\[ X=\frac{N}{M}, \] where $N$ is an $n-$variate
Gaussian vector and $M$ is a scalar Gaussian variable, independent
of $N$. Applying the above results, we deduce that they can be
obtained as gnomonic projections of uniformly distributed points on the
$n-$sphere, as illustrated in
the following figure in the case $n=3$: as point $P$ describes
uniformly the surface of $S_{3}$, point $M_{2}$ in the plane
$X_{3}=1$ is distributed with Tsallis law $T_{1/3,I_{2}}$.

\begin{center}
\begin{figure}
\includegraphics
[scale=0.7]{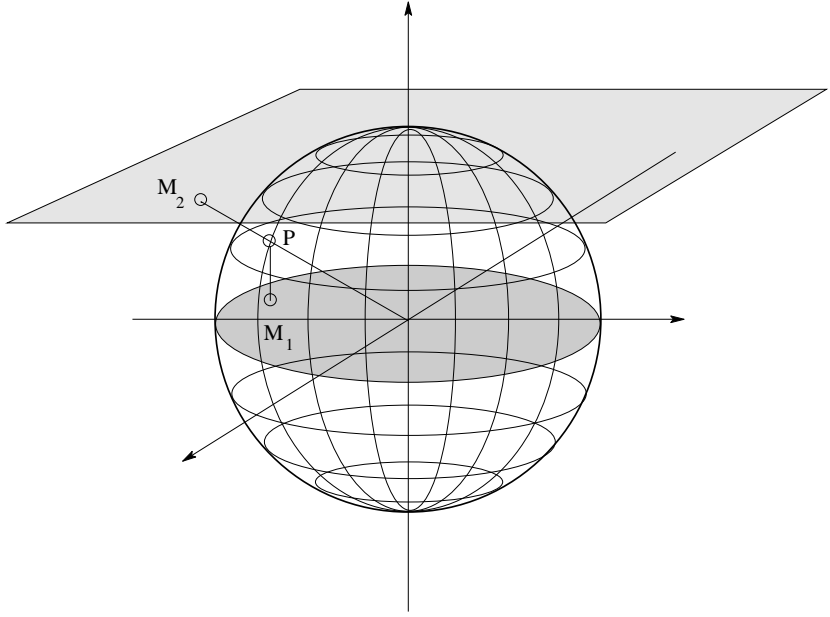}
\caption{ point $M_{2}$ distributed with Tsallis law $T_{1/3,I_{2}} $ while point $M_{1}$ is distributed with Tsallis law $T_{-1,I_{2}}$ }
\end{figure}
\end{center}

\section{Application: the ideal gas}

We adopt here notations of Minlos \cite{minlos}. Denote as $q_{i}$
the velocity of the i-th particle among $N$ of an ideal gas in
volume $\Lambda$. Let us define the set
\[\Omega_{\Lambda,N,E}=\left\{
\left(q_{1},\dots,q_{N}\right) \mid H\left(q_{1},\dots,q_{N}\right)=E\right\}
\] where the Hamiltonian writes\[H\left(q_{1},\dots,q_{N}\right)
=\sum_{i=1}^{N}\frac{1}{2}mq_{i}^{2}.\] In the microcanonical
setup, one assumes that the velocities have a distribution $\nu$
uniform on surface
$\Omega_{\Lambda,N,E}$:\[\nu\left(q_{1},\dots,q_{N}\right)
=\frac{2\vert\Lambda\vert^{N}\pi^{3N/2}}{m\Gamma
\left(\frac{3N}{2}\right)}\left(\frac{2E}{N}\right)^{\frac{3N}{2}-1}.\]
Let us now consider a subsystem of the gas, consisting of
$N_{0}<N$ particles in a volume $\Lambda_{0}\subset\Lambda$: their
distribution can be easily computed as\begin{equation}
\nu\left(q_{1},\dots,q_{N_{0}}\right)\propto\left(E-\sum_{i=1}^{N_{0}}\frac{1}{2}mq_{i}^{2}\right)_{+}^{\frac{3}{2}\left(N-N_{0}\right)-1}.\label{eq:gas}\end{equation}
A usual approach consists then in taking the thermodynamic limit:
assuming the total volume $\Lambda$ converges to $\mathbb{R}^{\mathnormal{3}}$, the
number of particle per volume
$\frac{N}{\vert\Lambda\vert}$ converges to density $\rho$ and the energy per volume $\frac{E}{\vert\Lambda\vert}$ converges to energy density $e$, we
deduce\[\nu\left(q_{1},\dots,q_{N_{0}}\right)
\propto\exp\left(-\frac{3\rho}{2e}\sum_{i=1}^{N_{0}}\frac{1}{2}mq_{i}^{2}\right)\]
and recover the celebrated Boltzmann distribution of velocities.

If, however, we do not take the thermodynamic limit but remain
under the finite-dimensional assumption, then we remark that
distribution (\ref{eq:gas}) is an $N_{0}-$variate Tsallis
distribution with nonextensivity index
\begin{equation}
q=\frac{N-N_{0}}{N-N_{0}-\frac{2}{3}}>1.\label{eq:q>1}\end{equation} In
other terms, the distribution of the finite dimensional ideal gas
in the microcanonical ensemble maximizes Tsallis entropy with
parameter $q$ as defined by (\ref{eq:q>1}). As $N-N_{0}$ is
exactly the dimension of the heat bath seen by the $N_{0}$
particles, we may conclude that, in this example, Tsallis entropy appears
as the finite-dimensional counterpart of Shannon entropy. This
idea was already derived by one of the authors \cite{Plastino}
years ago, but without the details we provide here. Moreover, Adib
et al. \cite{adib} have characterized Tsallis $q>1$ distributions as marginals
of uniform distributions, in the case of more general Hamiltonians
verifying an homogeneity property, but they assume the (stronger)
ergodic hypothesis to derive this result.

\end{document}